\documentclass[aps,showpacs,showkeys,superscriptaddress]{revtex4}

\usepackage{graphicx}
\usepackage{makeidx}
\usepackage[]{hyperref}

\usepackage{amsmath}
\usepackage{amssymb,epsfig,graphics}
\usepackage{amscd}
\usepackage{verbatim}

\usepackage{amsmath}
\usepackage{amsthm}
\usepackage{amsfonts}

\usepackage[font=large,labelfont=bf]{caption}

\begin{document}
\title{The chiral massive fermions in the graphitic wormhole}

\author{R. Pincak}\email{pincak@saske.sk}
\affiliation{Institute of Experimental Physics, Slovak Academy of Sciences,
Watsonova 47,043 53 Kosice, Slovak Republic}
\affiliation{Bogoliubov Laboratory of Theoretical Physics, Joint
Institute for Nuclear Research, 141980 Dubna, Moscow region, Russia}

\author{J. Smotlacha}\email{smota@centrum.cz}
\affiliation{Bogoliubov Laboratory of Theoretical Physics, Joint
Institute for Nuclear Research, 141980 Dubna, Moscow region, Russia}
\affiliation{Faculty of Nuclear Sciences and Physical Engineering, Czech Technical University, Brehova 7, 110 00 Prague,
Czech Republic}

\date{\today}



\begin{abstract}
The graphitic wormhole is in the focus of physical interest because of its interesting properties which can remotely resemble the concept of the space wormhole. Apart from the usual applications of the carbon nanostructures like the electronic computer devices, it seems to be a good material for the accumulation of the electric charge and different kinds of molecules, e.g. the hydrogen molecules, which enables using this material for the storage of the new kinds of fuel. Here, we present the geometric and electronic structure and calculate the zero modes of this material and its possibly significant derivate, the perturbed wormhole. Next, the influence of some additional factors on the electronic structure like the changes of the Fermi velocity close to the wormhole bridge as well as the spin-orbit interaction will be investigated.
On this basis, we predict the massive chiral fermions in the connecting wormhole nanotube.
\end{abstract}

\maketitle

\section{Introduction}\

During the last years, investigation of nanostructures experiences great development. They are good materials for constructing computer electronic devices, but their unique chemical and mechanical properties promise a wide application in many mutually different branches. The hexagonal carbon lattice structure and its variations are responsible for all of these extraordinary phenomena.

In the passed years, the basic forms of the nanostructures were investigated: the fullerene, graphene, nanotubes, nanohorns, nanocones, nanowires, etc.  Now, the steps are implemented for the preparation of more complicated forms like the wormhole \cite{herrero1, herrero2} and for the description of the electronic peculiarities of this type of structure. Due to the repulsion forces between the carbon atoms on the different graphene sheets, it seemed to be very difficult to pursue this problem on a different than the theoretical basis. However, the progress of the laboratory methods promises to get good results in this field. The presence of the wormholes in the synthesized structures could be proven by the detection of the zero energy modes of the fermions in the places of the wormhole bridges.

The electronic properties of the nanostructures are among others determined by their geometry, which is given by the presence of the defects \cite{tuzun, fiscal}. However, the deformation of the molecules can also be achieved mechanically or by the thermal influence. In the case of the former we speak about the so-called "straintronics" - the origin of the electronic structure comes from the mechanical strain affecting the molecule. This is exactly a new window to the world of nanostructures and devices similarly as was "spintronics" a decade ago. Moreover, Klein tunneling and cone transport in AA-stacked bilayer graphene give raise to the possibility of cone-tronic devices based on AA-stacked BLG \cite{cone}.

In the case of the wormhole, the geometry can also cause a shift of the Fermi energy at different distances from the wormhole center, which could direct the electron flux to this center, and in this way, the electric charge could be accumulated. The extraordinary deformation of the wormhole structures could highly influence the character of the Fermi velocity, relativistic effects could appear in these conditions. First of all, it could significantly change the mass of fermions which are usually considered to be massless. Another effect which could significantly influence the character of fermions is the spin-orbit coupling or interaction (SOC) for the fermions located in the carbon nanotube. We will show that the strength of this interaction strongly depends on the radius of the nanotube, it should be inversely proportional to this radius. As a result, the chiral (massive) fermions should be detected around the wormhole bridge.

This paper is an extension of work \cite{pinw}. Here, we investigate the graphitic wormhole: we calculate its electronic structure and zero modes and look through the corrections coming from the relativistic effects and the spin-orbit interaction. Similar calculations of the electronic structure and the zero modes will be carried out for the perturbed wormhole.  Furthermore, we will predict the effect of the graphene blackhole for this case.\\

\section{Continuum gauge field-theory}\

The electronic structure can be characterized by the local density of states ($LDoS$) which gives the number of the electronic states per unit interval of energies and per the unit area of the molecular surface. The reason is an intuitive assumption that there exists a direct connection between this quantity and the electronic conductivity. For its calculation, in the case of the wormhole and the perturbed wormhole, we use the continuum gauge field-theory.

In the continuum gauge field-theory, we consider the continuum approach of the carbon lattice \cite{mele}. So at each point of the molecular surface we take into the account the influence of different gauge fields and we insert them into the Dirac-like equation for the electron. It has the form
\begin{equation}\label{Dirac}{\rm i}v_F\sigma^{\mu}[\partial_{\mu}+\Omega_{\mu}-ia_{\mu}-ia_{\mu}^W-iA_{\mu}]\psi=E\psi,\end{equation}
in which the different terms have the meaning of different kinds of properties. The matrices $\sigma^{\alpha}$ which represent the $\gamma$ matrices, which are used for the $4$-component case, have the form of the Pauli matrices. The Fermi velocity $v_F$, the spin connection \begin{equation}\Omega_{\mu}=\frac{1}{8}\omega^{\alpha\beta}_{\mu}[\sigma_{\alpha},\sigma_{\beta}],\end{equation}
and the covariant derivative $\nabla_{\mu}=\partial_{\mu}+\Omega_{\mu}$ are connected with the metric. Next, the gauge fields
$a_{\mu}, a_{\mu}^W$ ensure the continuation of the wave function with respect to the angular coordinate, and they are caused by the presence of the defects and by the rotational symmetry, respectively. The gauge field $A_{\mu}$ characterizes the possible magnetic field.

The wave function has two components:
\begin{equation}\psi=\left(\begin{array}{c}\psi_A \\ \psi_B\end{array}\right),\end{equation}
where $A,B$ correspond to inequivalent sublattices of the hexagonal plane. The solution is found by using the substitution
\begin{equation}\psi=\left(\begin{array}{c}\psi_A \\ \psi_B\end{array}\right)=\frac{1}{\sqrt[4]{g_{\varphi\varphi}}}\left(\begin{array}{c}u_j(\xi)e^{i\varphi
j}\\ v_j(\xi)e^{i\varphi(j+1)}\\\end{array}\right),\hspace{2.5mm} j=0,\pm
1,...,\end{equation}
so
\begin{equation}\label{system}\frac{\partial_{\xi}u_j}{\sqrt{g_{\xi\xi}}}-\frac{\widetilde{j}}
{\sqrt{g_{\varphi\varphi}}}u_j
=Ev_j,\hspace{1cm}-\frac{\partial_{\xi}v_j}{\sqrt{g_{\xi\xi}}}-\frac{\widetilde{j}}
{\sqrt{g_{\varphi\varphi}}}v_j
=Eu_j,\end{equation}
where
\begin{equation}\widetilde{j}=j+1/2-a_{\varphi}-a_{\varphi}^W-A_{\varphi},\end{equation}
$g_{\xi\xi}, g_{\varphi\varphi}$ are the metric coefficients and $j$ is the angular momentum. Then, if $u, v$ are the normalized solutions, the local density of states can be calculated as
\begin{equation}
LDoS(E,\xi_0)=u^2(E,\xi_0)+v^2(E,\xi_0).
\end{equation}\\

\section{The Graphitic Wormhole}\

\begin{figure}[htbp]
\includegraphics[width=70mm]{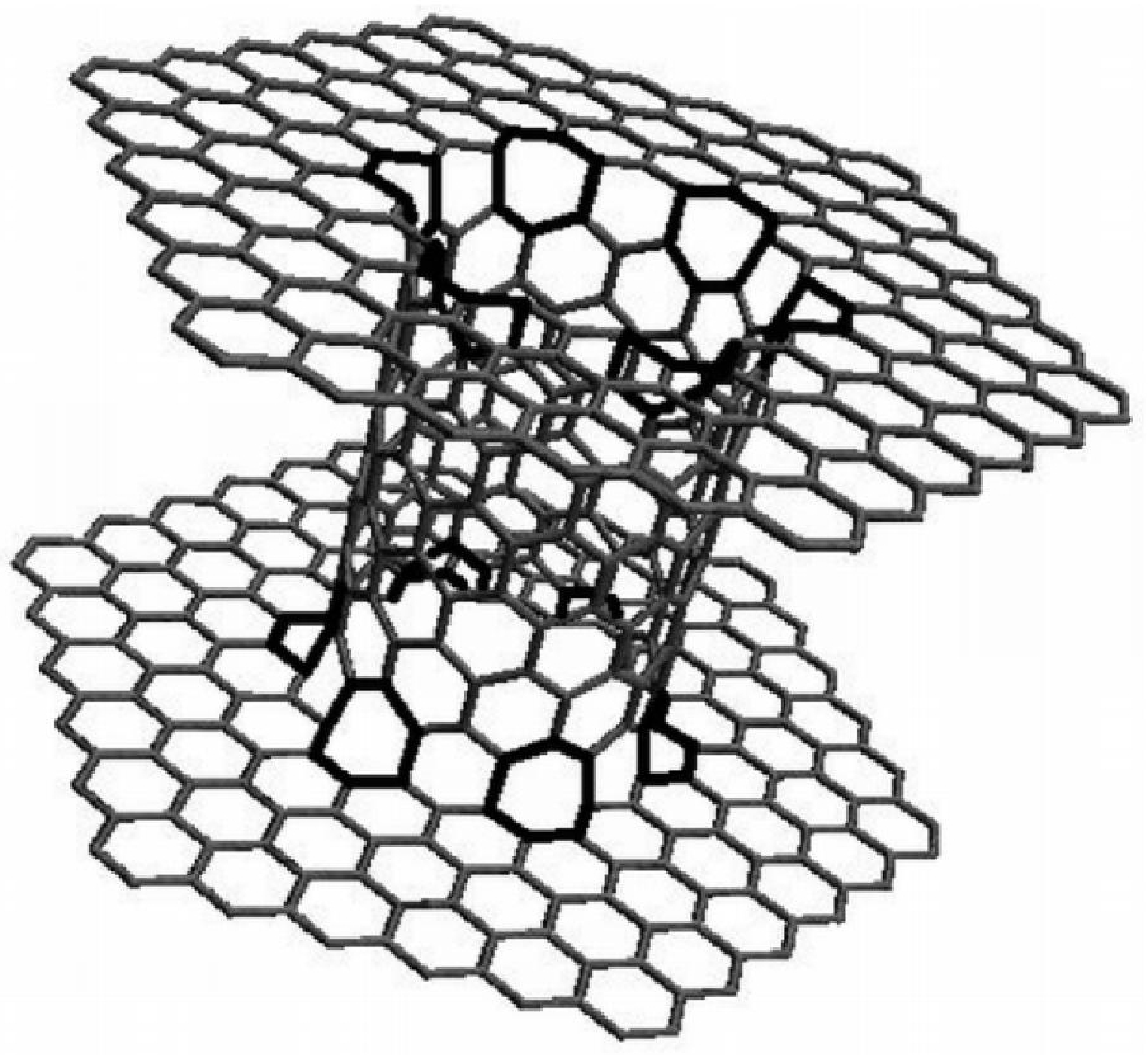}
\caption{{\fontfamily{phv}\selectfont {\fontsize{11}{0}\selectfont \textbf{Wormhole structure.}}}}\label{fgWorm}
\end{figure}

The wormhole (Fig. \ref{fgWorm}) is usually understood as a form which arises when two graphene sheets are connected together with the help of the connecting nanotube \cite{herrero1, herrero2}. This can be achieved by the supply of 2 sets of 6 heptagonal defects onto both sides of the given nanotube. Of course, this brings the restrictions on the form of the nanotube - the chirality must be $(6n, 6n)$ armchair or $(6n, 0)$ zig-zag. Furthermore, because of the physical limitations, the radius of the nanotube must be much larger than its length.\\

The metric tensor of the wormhole has the form
\begin{equation}\label{metric}g_{\mu\nu}=\Lambda^2(r_{\pm})\left(\begin{array}{cc}1 & 0\\0 & r_{\pm}^2\end{array}\right),\hspace{5mm}
\Lambda(r_{\pm})=\left(a/r_{\pm}\right)^2\theta (a-r_{\pm})+\theta (r_{\pm}-a),
\end{equation}
where $\theta$ is the Heaviside step function, $r_-,r_+$ are the polar coordinates corresponding to the lower and upper graphene sheet, respectively and $a=\sqrt{r_-r_+}$ is the radius of the wormhole.\\

\subsection{Electronic structure}\

\begin{figure}[htbp]
\includegraphics[width=90mm]{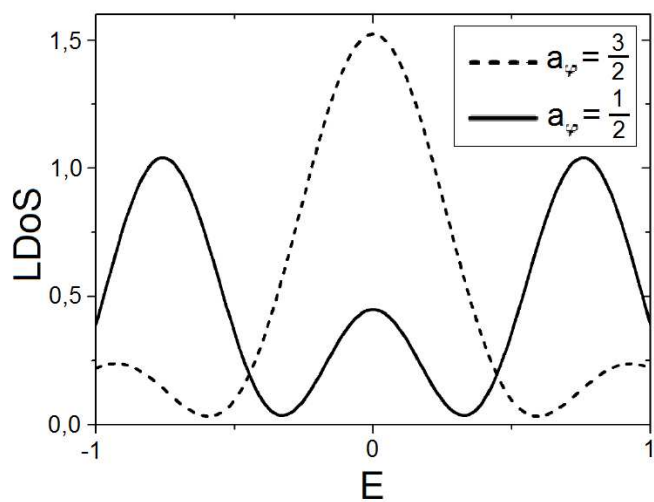}
\caption{{\fontfamily{phv}\selectfont {\fontsize{11}{0}\selectfont \textbf{Local density of states on the bridge of the graphitic wormhole for different values of $a_{\varphi}$.}}}}\label{fg1}
\end{figure}

Now the solution of the Dirac equation will be found for the case of the wormhole. In this case, we use the form of the metric used in (\ref{metric}). Next, the effective flux caused by the presence of the defects is included in the gauge field $a_{\mu}$ and for the particular polar components it has the values
\begin{equation}a_{\varphi}=\frac{3}{2},\hspace{1cm}a_{r}=0\end{equation}
for these two possibilities: the first corresponds to the case when the chiral vector has the form $(6n,6n)$, the second corresponds to the case when the chiral vector has the form $(6n,0)$ with $n$ divisible by $3$. In the case of the chiral vector of the form $(6n,0)$, where $n$ is not divisible by $3$, the components of the corresponding gauge field have the form
\begin{equation}a_{\varphi}=\frac{1}{2},\hspace{1cm}a_{r}=0.\end{equation}
Regarding that the components of the spin connection are
\begin{equation}\Omega_{\varphi}=-\frac{{\rm i}}{2}\sigma_3\left(r\frac{\Lambda'(r)}{\Lambda(r)}+1\right),\hspace{1cm}\Omega_r=0,\end{equation}
and after the substitution into (\ref{Dirac}) we get the equation
\begin{equation}{\rm i}v_F\sigma^{\mu}(\partial_{\mu}+\Omega_{\mu}\mp{\rm i}\,a_{\mu})\psi^{\pm}=\varepsilon\psi^{\pm},\end{equation}
where each sign corresponds to a different Dirac point. Concretely,
\begin{equation}\label{geq}-{\rm i}v_F\left(\partial_r+\frac{1}{r}{\rm i}\partial_{\theta}\mp \frac{a_{\varphi}}{r}+\frac{1}{2r}\right)\psi_B^{\pm}=\varepsilon\psi_A^{\pm},\hspace{1cm} -{\rm i}v_F\left(\partial_r-\frac{1}{r}{\rm i}\partial_{\theta}\pm \frac{a_{\varphi}}{r}+\frac{1}{2r}\right)\psi_A^{\pm}=\varepsilon\psi_B^{\pm}\end{equation}
for $r\geq a$ and
\begin{equation}\label{leq}{\rm i}v_F\left(\frac{r}{a}\right)^2\left(\partial_r-\frac{1}{r}{\rm i}\partial_{\theta}\pm \frac{a_{\varphi}}{r}-\frac{1}{2r}\right)\psi_B^{\pm}=\varepsilon\psi_A^{\pm},\hspace{1cm} {\rm i}v_F\left(\frac{r}{a}\right)^2\left(\partial_r+\frac{1}{r}{\rm i}\partial_{\theta}\mp \frac{a_{\varphi}}{r}-\frac{1}{2r}\right)\psi_A^{\pm}=\varepsilon\psi_B^{\pm}\end{equation}
for $0<r\leq a$. For $r\geq a$, the solution is
\begin{equation}\psi^{\pm}=\left(\begin{array}{c}\psi^{\pm}_A(r,\varphi) \\ \psi^{\pm}_B(r,\varphi)\end{array}\right)=
c_1\left(\begin{array}{c}J_{n\mp a_{\varphi}-1/2}(kr) \\ -{\rm i\,sgn}\,\varepsilon J_{n\mp a_{\varphi}+1/2}(kr)\end{array}\right)+
c_2\left(\begin{array}{c}Y_{n\mp a_{\varphi}-1/2}(kr) \\ -{\rm i\,sgn}\,\varepsilon Y_{n\mp a_{\varphi}+1/2}(kr)\end{array}\right),
\end{equation}
where $J_n(x)$ and $Y_n(x)$ are the Bessel functions and the energy $\varepsilon=\pm v_Fk$. The local density of states for different values of the component of the gauge field $a_{\varphi}$ is seen in Fig. \ref{fg1}.\\

\begin{figure}[htbp]
\includegraphics[width=95mm]{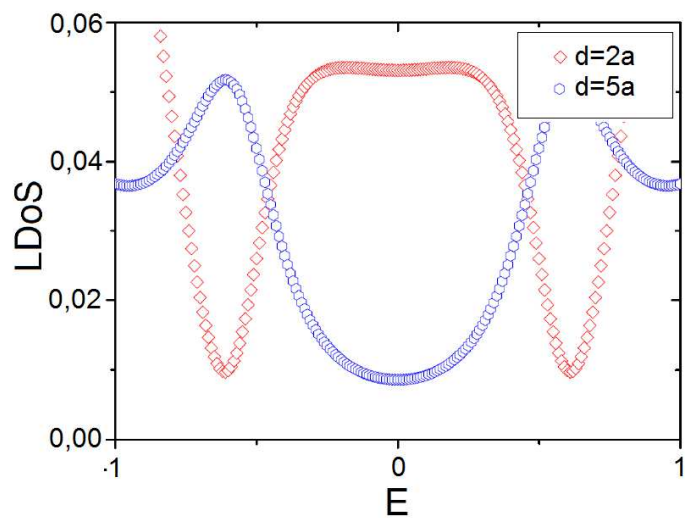}
\caption{{\fontfamily{phv}\selectfont {\fontsize{11}{0}\selectfont \textbf{Local density of states at the distances twice and five times the radius of the wormhole from the wormhole bridge.}}}}\label{fgDV}
\end{figure}

In Fig. \ref{fgDV}, we can see the local density of states at different distances from the wormhole bridge.\\

\subsection{Zero modes}\

The zero modes solve the Dirac equation for the zero energy. If we choose the component $\psi_A^{\pm}$ of the solution to be equal to zero, we get from (\ref{geq}), (\ref{leq})
\begin{equation}\left(\partial_r-\frac{1}{r}{\rm i}\partial_{\theta}\mp \frac{a_{\varphi}}{r}+\frac{1}{2r}\right)\psi_B^{\pm}=0\end{equation}
for $r\geq a$ and
\begin{equation}\left(\partial_r-\frac{1}{r}{\rm i}\partial_{\theta}\pm \frac{a_{\varphi}}{r}-\frac{1}{2r}\right)\psi_B^{\pm}=0\end{equation}
for $0<r\leq a$. For $\psi_B^-$ and the value $a_{\varphi}=\frac{3}{2}$, it has the solution
\begin{equation}\label{zero1}\psi_B^-(r,\varphi)\sim r^{-n-2}e^{{\rm i}\,n\varphi}\end{equation}
for $r\geq a$ and
\begin{equation}\label{zero2}\psi_B^-(r,\varphi)\sim r^{-n+2}e^{{\rm i}\,n\varphi}\end{equation}
for $0<r\leq a$. It is both strictly normalizable only for $n=0$, so this is the only solution. In a similar way, we can calculate the zero modes for the component $\psi_B^+$. We get analogical solution for $\psi_A^{\pm}$ if we choose the component $\psi_B^{\pm}$ of the solution to be equal to zero.

For the value $a_{\varphi}=\frac{1}{2}$, possible solutions are not strictly normalizable. So the zero modes exist only for the case of the connecting nanotube being armchair or zig-zag with the chiral vector $(6n,0)$, $n$ divisible by $3$. In other cases, the zero modes do not exist.\\

\begin{figure}[htbp]
\includegraphics[width=150mm]{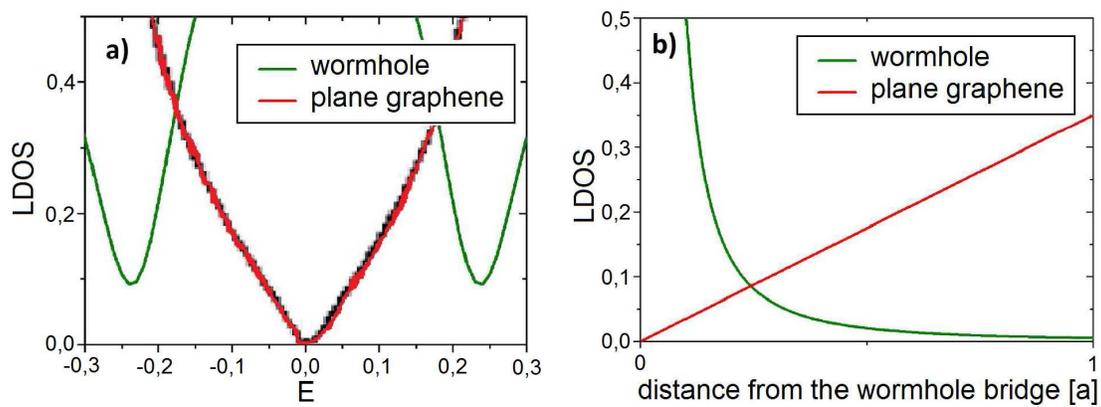}
\caption{{\fontfamily{phv}\selectfont {\fontsize{11}{0}\selectfont \textbf{Comparison of the properties of the wormhole and the plane graphene: a) local density of states, b) zero modes.}}}}\label{fg2}
\end{figure}

Figure \ref{fg2}a  shows very big localization of the $LDoS$ near Fermi energy on the bridge of the wormhole (in comparison with the $LDoS$ of the plane graphene). This big localization or singularity for zero mode solutions could be experimentally observed. In Fig. \ref{fg2}b, the comparison of the zero modes of the wormhole and the graphene is plotted.

Recently, in work \cite{pinbil} some peculiarities in the bilayer graphene were analytically predicted.
A possible indication of the wormhole could be found in \cite{bilayer, trilayer}, where a new type of zero modes is investigated. These zero modes could be the zero modes studied in this subsection applied to the case of the smallest wormhole.\\

\subsection{Case of massive fermions}\

In the continuum gauge field-theory, we suppose that the fermions appearing in the Dirac equation have the zero mass, or more precisely, the mass is very small in comparison with the energy. But it was shown in \cite{pin,vf} that the Fermi velocity changes and needs to be renormalized due to the elasticity and the deformations in graphene. In the investigated case of the graphitic wormhole, which is very specific because of the big deformations, the velocity of the fermions close to the wormhole bridge could achieve such values that the relativistic effects can appear or break off the symmetry \cite{chenwei}. The result is that the mass of fermions would be non-negligible. (The same situation is also for the graphene bilayer where especially massive particles are predicted, see \cite{blvf0, blvf}.) Moreover, the radius of the wormhole and its bridge is very small in comparison with the size of the upper and the lower graphene sheet, and by folding the sheet into a tube they acquire nonzero effective mass as they move along the tube axis. This change of the space topology of graphene from 2D to 1D space compactification is similar to the string theory compactification. It means that we can image a wormhole connecting nanotubes as the 1D object.

\begin{figure}[htbp]
\includegraphics[width=150mm]{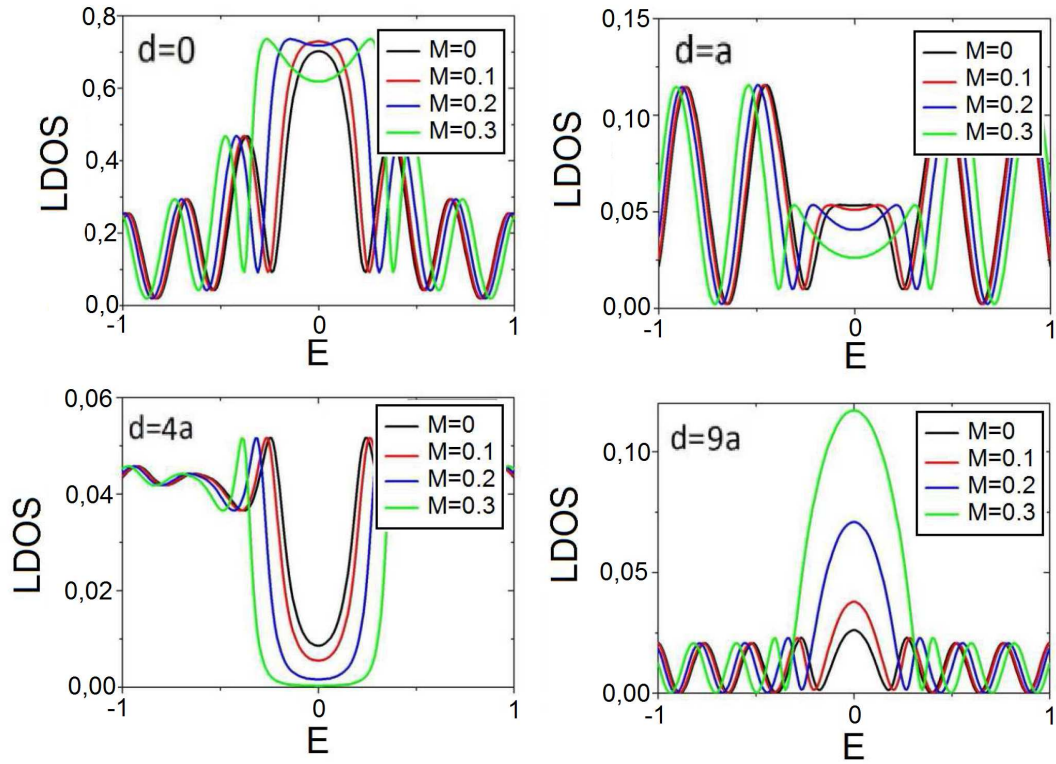}
\caption{{\fontfamily{phv}\selectfont {\fontsize{11}{0}\selectfont \textbf{Comparison of $LDoS$ for different masses of fermions at different distances $d$ from the wormhole bridge.}}}}\label{fgMass}
\end{figure}

This indicates a necessity to incorporate a term containing the value of the mass into the Dirac equation (\ref{Dirac}). For the purpose of the derivation of the possible solution of this question, we go through the system of equations (\ref{system}), which can be transformed into the differential equation of the second order
\begin{equation}\label{secord}\left(\partial_{\xi\xi}-\frac{1}{2g_{\xi\xi}}\partial_{\xi}g_{\xi\xi}+\frac{\tilde{j}}{2}\sqrt{\frac{g_{\xi\xi}}
{g_{\varphi\varphi}^3}}\partial_{\xi}g_{\varphi\varphi}-\tilde{j}^2\frac{g_{\xi\xi}}{g_{\varphi\varphi}}+E^2g_{\xi\xi}\right)u_j=0.
\end{equation}
For the purpose of simplification, we will suppose the cylindrical geometry, i.e. the radius vector of the point at the surface will have the form
\begin{equation}\vec{R}=(R\cos\varphi, R\sin\varphi, \xi),\end{equation}
where $R$ is the radius of the cylinder. In this case, the form of equation (\ref{secord}) will be changed into
\begin{equation}\left(\partial_{\xi\xi}+E^2-\frac{\tilde{j}^2}{R^2}\right)u_j=0.\end{equation}
The solution of this equation has the form \cite{tubmass}
\begin{equation}u_j(\xi)=Ae^{k\xi}+Be^{-k\xi},\end{equation}
where
\begin{equation}k=\sqrt{\frac{\tilde{j}^2}{R^2}-E^2}.\end{equation}
From \cite{ten} follows that a similar form has the dispersion relation associated with the massive 1D Dirac equation:
\begin{equation}k=\sqrt{M^2-E^2},\end{equation}
where $M$ is the mass of the corresponding fermion. It is proven in \cite{tubmass} that indeed, for a suitable choice of the parameters, the 2D massless case is analogous to the 1D massive case. This serves as an impulse to rewrite equation (\ref{secord}) in the form
\begin{equation}\label{massDir}\left(\partial_{\xi\xi}-\frac{1}{2g_{\xi\xi}}\partial_{\xi}g_{\xi\xi}+\frac{\tilde{j}}{2}\sqrt{\frac{g_{\xi\xi}}
{g_{\varphi\varphi}^3}}\partial_{\xi}g_{\varphi\varphi}-\tilde{j}^2\frac{g_{\xi\xi}}{g_{\varphi\varphi}}+(E^2-M^2)g_{\xi\xi}\right)u_j=0,
\end{equation}
where $M$ is the mass of the corresponding fermion. Then, for different values of $M$, we can find the corrections of the $LDoS$ for the graphitic wormhole. For different distances from the wormhole bridge, we can see these corrections in Fig. \ref{fgMass}. We predict that these massive particles arising in the wormhole nanotubes could create energy bulks on the wormhole bridge and near the wormhole bridge which should be experimentally measured by the STM or Raman spectroscopy \cite{enerbulk}. Another possibility to identify the wormhole structure comes from the fact that the massive particles could create strain solitons and topological defects on the bridge of the bilayer graphene which should propagate throughout the graphene sheet. These are almost macroscopic effects and should be caught by the experimentalists \cite{soliton}.\\

\subsection{Spin-orbit coupling in the wormhole connecting nanotube}\

An important measurable quantity in the carbon nanostructures (including the nanotubular part of the graphitic wormhole) is the spin-orbit coupling (SOC) \cite{spoc1, spoc2}. Considering this influence, the 2-component Dirac equation is changed into the usual 4-component form. As a consequence, the chiral fermions should be detected close to the wormhole bridge. We will show that the smaller is the radius of the wormhole bridge, the stronger this effect should be.\\

We have 2 sources of the SOC: the first, the interatomic one that preserves the $z$-component of the spin and the second, the so-called Rashba-type coming from the external electric field, which conserves the $z$-component of the angular momentum $J_z$. In both cases, the strength of the SOC is influenced by the nonzero curvature. Here, we will be interested in the first, interatomic source of the SOC.


Considering the SOC we can write the Dirac equation for the nanotube in the form
\begin{equation}\hat{H}\left(\begin{array}{c}F_A^K \\ F_B^K\end{array}\right)=\left(\begin{array}{cc}0 & \hat{f}\\ \hat{f}^{\dagger} & 0\end{array}\right)
\left(\begin{array}{c}F_A^K \\ F_B^K\end{array}\right)=E\left(\begin{array}{c}F_A^K \\ F_B^K\end{array}\right),\end{equation}
where
\begin{equation}F_A^K=\left(\begin{array}{c}F_{A,\uparrow}^K \\ F_{A,\downarrow}^K\end{array}\right),\hspace{5mm}
F_B^K=\left(\begin{array}{c}F_{B,\uparrow}^K \\ F_{B,\downarrow}^K\end{array}\right).\end{equation}
The expression $\hat{f}$ has the form
\begin{equation}\hat{f}=\gamma(\hat{k}_x-i\hat{k}_y)+i\frac{\delta\gamma'}{4R}\hat{\sigma}_x(\vec{r})-\frac{2\delta\gamma p}{R}\hat{\sigma}_y,\end{equation}
where
\begin{equation}\hat{k}_x=-i\frac{\partial}{R\partial\theta},\hspace{5mm}\hat{k}_y=-i\frac{\partial}{\partial y},\hspace{5mm}
\hat{\sigma}_x(\vec{r})=\hat{\sigma}_x\cos\theta-\hat{\sigma}_z\sin\theta.\end{equation}
Next,
\begin{equation}\gamma=-\frac{\sqrt{3}}{2}aV^{\pi}_{pp},\hspace{5mm}\gamma'=-\frac{\sqrt{3}}{2}a(V^{\sigma}_{pp}-V^{\pi}_{pp}),
\hspace{5mm}p=1-\frac{3\gamma'}{8\gamma},\end{equation}
$a$ being the length of the atomic bond, $V^{\sigma}_{pp}, V^{\pi}_{pp}$ are the hopping integrals for the $\sigma$ and $\pi$ bond, respectively.

For the interatomic source of the SOC we have
\begin{equation}\delta=\frac{\Delta}{3\epsilon_{\pi\sigma}},\hspace{5mm}\Delta=i\frac{3\hbar}{4m^2c^2}
\langle x_l|\frac{\partial V}{\partial x}\hat{p}_y-\frac{\partial V}{\partial y}\hat{p}_x|y_l\rangle\end{equation}
with the difference of the energies of the relevant $\pi$ and $\sigma$ orbitals
\begin{equation}\epsilon_{\pi\sigma}=\epsilon^{\pi}_{2p}-\epsilon^{\sigma}_{2p},\end{equation}
$x_l$ and $y_l$ are the local coordinates. By applying the transformation
\begin{equation}\hat{H}'=\hat{U}\hat{H}\hat{U}^{-1}\end{equation}
with
\begin{equation}\hat{U}=\left(\begin{array}{cc}exp(i\hat{\sigma_y}\frac{\theta}{2}) & 0\\ 0 & exp(i\hat{\sigma_y}\frac{\theta}{2})\end{array}\right)\end{equation}
the transformed Hamiltonian $\hat{H}'$ will has the form
\begin{equation}\hat{H}'=\hat{H}_{kin}+\hat{H}_{SOC},\end{equation}
\begin{equation}\hat{H}_{kin}=-i\gamma\left(\partial_y{\rm Id}_2\otimes\hat{s}_y+\frac{1}{R}\partial_{\theta}{\rm Id}_2\otimes\hat{s}_x\right),\end{equation}
\begin{equation}\hat{H}_{SOC}=\lambda_y\hat{\sigma}_x\otimes\hat{s}_y-\lambda_x\hat{\sigma}_y\otimes\hat{s}_x,\end{equation}
where $\hat{H}_{SOC}$ corresponds to the spin-orbit coupling. The operators $\hat{s}_{x,y,z}$ are the Pauli matrices, which transform the wave function of the $A$ sublattice into the wave function of the $B$ sublattice and vice versa.

In our model, the SOC is induced by the curvature and is described with
the aid of two strength parameters $\lambda_x$ and $\lambda_y$ which have, in
the case of the single wall carbon nanotube with different magnitude, the form
\begin{equation}\lambda_x=\frac{\gamma}{R}\left(\frac{1}{2}+2\delta p\right),\hspace{5mm}\lambda_y=-\frac{\delta\gamma'}{4R}.\end{equation}
Here, $|\lambda_y|\ll|\lambda_x|$ and for $R\rightarrow 0$, both strengths go to infinity, as we required.\\

So reminding the results of the previous section, the chiral massive fermions should be detected around the wormhole bridge. The presented strength constants $\lambda_x, \lambda_y$ for the SOC as well as the Dirac-like equation (\ref{massDir}) for the massive fermions have not yet been published anywhere. \\

More complicated forms can arise: the nanotube in the wormhole center can be perturbed. Then, the geometry of the corresponding graphene sheets will be curved and this brings a significant change of the physical properties. We speak about the perturbed wormhole.\\

\section{Case of perturbed wormhole}\

The wormhole is composed of 2 different kinds of nanostructures: the graphene and the nanotube. These two parts are connected with the help of 12 heptagonal defects. However, we can ask a question what happens if the number of the defects varies between 0 and 12. Then, the cases of 0 defects (the nanotube) and 12 defects (the wormhole) will be the limiting possibilities. The structure that arises in other cases will be called the perturbed wormhole.\\

\begin{figure}[htbp]
\includegraphics[width=160mm]{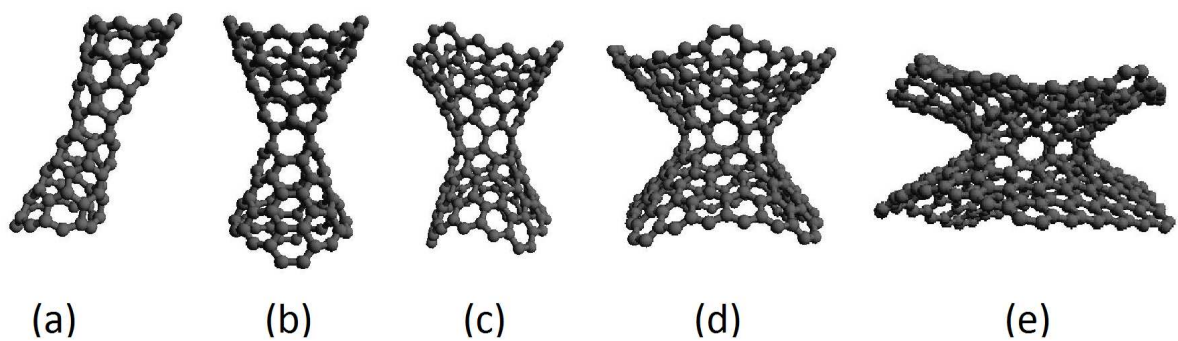}
\caption{{\fontfamily{phv}\selectfont {\fontsize{11}{0}\selectfont \textbf{Different forms of the perturbed wormhole: (a) 2 defects, (b) 4 defects, (c) 6 defects, (d) 8 defects, (e) 10 defects.}}}}\label{fgwor}
\end{figure}

\begin{figure}[htbp]
\includegraphics[width=80mm]{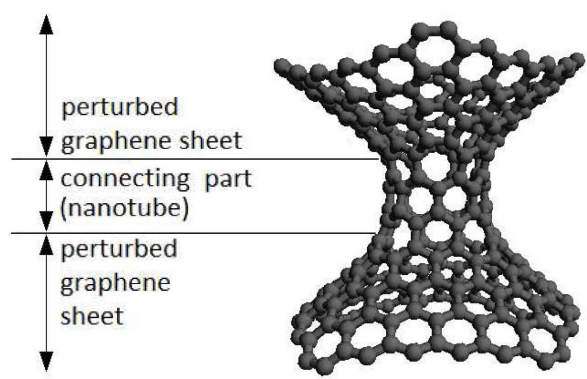}
\caption{{\fontfamily{phv}\selectfont {\fontsize{11}{0}\selectfont \textbf{The composition of the perturbed wormhole.}}}}\label{compworm}
\end{figure}

Possible forms of the perturbed wormhole (with the defects located in the middle) can be seen in Fig. \ref{fgwor}. Due to symmetry preservation, we consider only the even numbers of the defects, i.e. 2, 4, 6, 8 or 10. The defects can be located in the middle as well as at the edge of the connecting structure - in the former, we can say that the hight of the connecting nanotube is negligible (similarly to the case of the above draught structure of the wormhole containing 12 defects). We will be concerned with the case when the defects are located at the edge of the connecting nanotube. For this case, the composition of the perturbed wormhole is depicted in Fig. \ref{compworm}: it consists of the unperturbed connecting nanotube and 2 perturbed graphene sheets. The metric of the sheets does not depend on the length of the nanotube. That is why we will investigate the electronic structure of the sheets separately.\\

\begin{figure}[htbp]
\includegraphics[width=90mm]{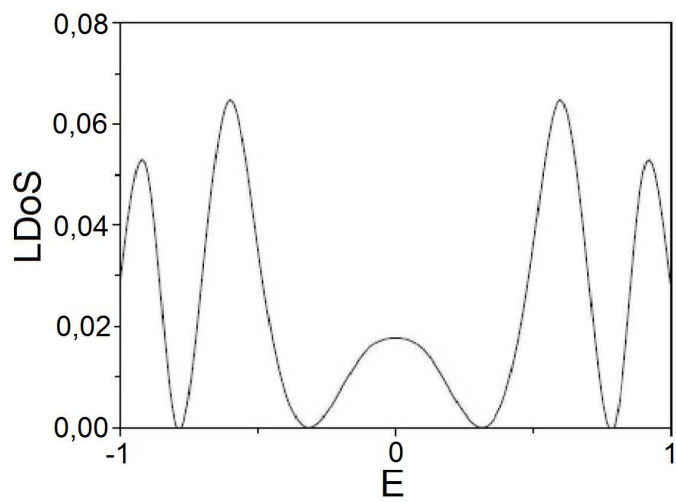}
\caption{{\fontfamily{phv}\selectfont {\fontsize{11}{0}\selectfont \textbf{Local density of states on the bridge of the graphitic perturbed wormhole.}}}}\label{fg3}
\end{figure}

The metric of the sheets can be draught by the radius vector
\begin{equation}\label{lvector}\overrightarrow{R}(z,\varphi)=\left(a\sqrt{1+\triangle z^2}\cos\varphi,
a\sqrt{1+\triangle z^2}\sin\varphi,z\right),
\end{equation}
where $\triangle$ is a positive real parameter; its value is derived from the number of the defects of the wormhole. In the case of $N=2$ defects, we can say that the value of this parameter is negligible, so $\triangle<<1$. Then, the nonzero components of the metric are
\begin{equation}g_{zz}=1+\frac{a^2\triangle^2 z^2}{1+\triangle z^2}\sim 1+a^2\triangle^2 z^2,\hspace{1cm}
g_{\varphi\varphi}=a^2(1+\triangle z^2).
\end{equation}
Next, we include into the calculations the nonzero components of the gauge fields:
\begin{equation}a_{\varphi}=N/4,\hspace{1cm}a_{\varphi}^W=-(2m+n)/3,\end{equation}
where $(n,m)$ is the chiral vector of the middle part of the connecting nanostructure. Then, regarding the form of the spin connection and by substitution in (\ref{Dirac}) we get the solution
\begin{equation}\psi_A(z)=C_{\triangle 1}D_{\nu_1}(\xi(z))e^{{\rm i}n\varphi}+C_{\triangle 2}D_{\nu_2}(i\xi(z)))e^{{\rm i}n\varphi},\end{equation}
\begin{equation}\psi_B(z)=\frac{C_{\triangle 1}}{E}\left(\partial_zD_{\nu_1}(\xi(z))-\frac{\widetilde{j}D_{\nu_1}(\xi(z))}{a}
(1-\frac{1}{2}\triangle^2z^2)\right)e^{-{\rm i}n\varphi}+
\frac{C_{\triangle 2}}{E}\left(\partial_zD_{\nu_2}(i\xi(z))-\frac{\widetilde{j}D_{\nu_2}(i\xi(z))}{a}
(1-\frac{1}{2}\triangle^2z^2)\right)e^{-{\rm i}n\varphi},\end{equation}
where
\begin{equation}\nu_1=i\frac{a^2\triangle-4 a^2 E^2+4ia\sqrt{\triangle}\widetilde{j}+4\widetilde{j}^2}{8a\sqrt{\triangle} \widetilde{j}},
\hspace{1cm}\nu_2=-i\frac{a^2\triangle-4 a^2 E^2-4ia\sqrt{\triangle}\widetilde{j}+4\widetilde{j}^2}{8a\sqrt{\triangle} \widetilde{j}},
\end{equation}
\begin{equation}\xi(z)= (-\triangle)^{1/4}\left(\sqrt{\frac{a}{
 2\widetilde{j}}} + \sqrt{\frac{2\widetilde{j}}{a}}z\right),
\end{equation}
$D_{\nu}(\xi)$ being the parabolic cylinder function. The functions $C_{\triangle 1}=
C_{\triangle 1}(E),\,C_{\triangle 2}=C_{\triangle 2}(E)$ serve as the normalization constants. We see the graph of the local density of states in Fig. \ref{fg3}.\\

\begin{figure}[htbp]
\includegraphics[width=60mm]{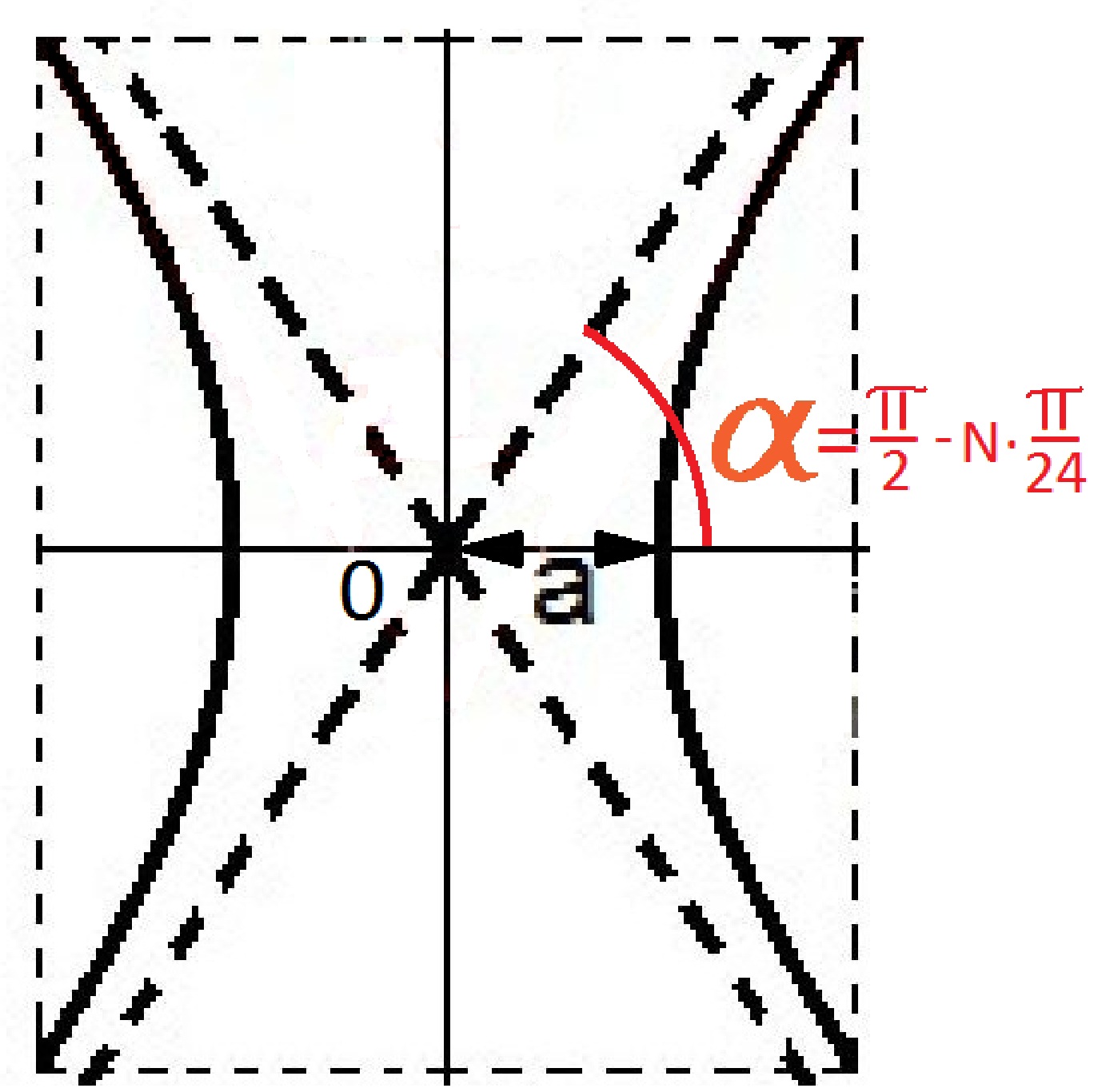}
\caption{{\fontfamily{phv}\selectfont {\fontsize{11}{0}\selectfont \textbf{Derivation of the $\triangle$ parameter.}}}}\label{fgTriangle}
\end{figure}

In the case of more than 2 defects, the value of $\triangle$ is non-negligible and we can get only numerical approximation of the $LDoS$. The derivation of the value of the parameter $\triangle$ follows from Fig. \ref{fgTriangle}.\\

In the middle section, the upper branch of the graphene sheet converges to the line $z=x\cdot\tan\alpha$, where we can suppose that the angle $\alpha$ depends on the number of the defects $N$ linearly, i.e. $\alpha=\frac{\pi}{2}-N\cdot\frac{\pi}{24}$. (In this case, $\alpha=\frac{\pi}{2}$ corresponds to 0 defects and $\alpha=0$ corresponds to 12 defects.) Simultaneously, from (\ref{lvector}) follows that asymptotically we have
\begin{equation}\overrightarrow{R}(z\rightarrow\infty,\varphi)\rightarrow\left(a\sqrt{\triangle}z\cos\varphi,
a\sqrt{\triangle}z\sin\varphi,z\right),\end{equation}
from which follows
\begin{equation}z=x\cdot\tan\alpha=\left(a\sqrt{\triangle}\right)^{-1}x,\end{equation}
so
\begin{equation}\triangle=\frac{1}{a^2\tan^2\alpha}=\frac{1}{a^2\tan^2\left(\frac{\pi}{2}-N\cdot\frac{\pi}{24}\right)}.\end{equation}

\begin{figure}[htbp]
\includegraphics[width=150mm]{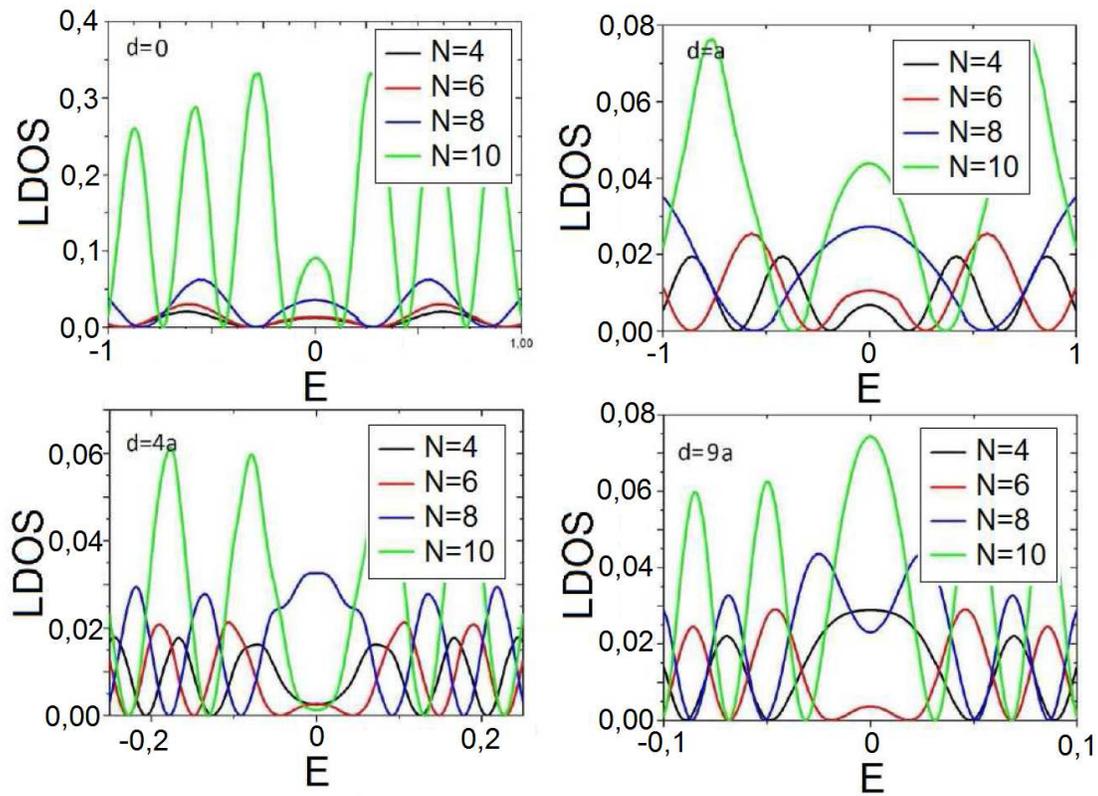}
\caption{{\fontfamily{phv}\selectfont {\fontsize{11}{0}\selectfont \textbf{Comparison of $LDoS$ for different numbers of the defects in the perturbed wormhole at different distances $d$ from the wormhole bridge.}}}}\label{pertvar}
\end{figure}

In Fig. \ref{pertvar}, we see the comparison of the $LDoS$ for different kinds of the perturbed wormhole. From the plots follows that the intensity is rising with the increasing number of the defects and it is closer and closer approaching the results in Fig. \ref{fg1}, where the case of 12 defects is shown. The localized states for zero energy is also in coincidence with the results presented for the bilayer and trilayer in \cite{bilayer, trilayer}.

\begin{figure}[htbp]
\includegraphics[width=100mm]{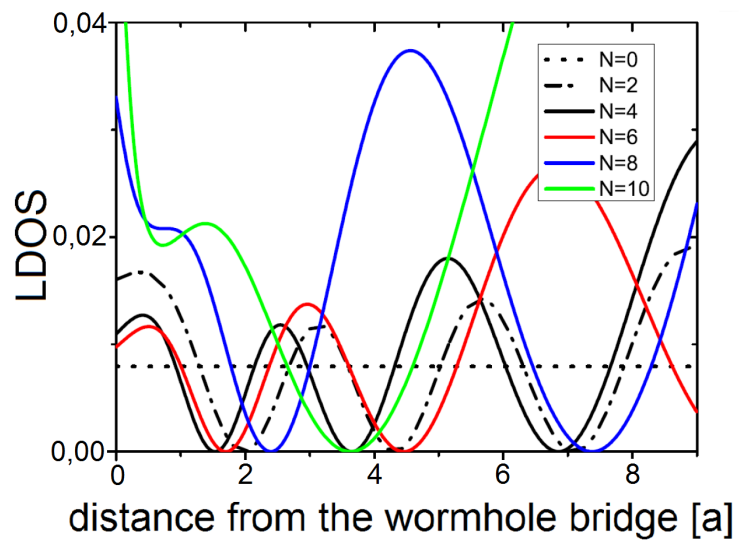}
\caption{{\fontfamily{phv}\selectfont {\fontsize{11}{0}\selectfont \textbf{Zero modes of the perturbed wormhole for different numbers of the defects.}}}}\label{zeropvar}
\end{figure}

In Fig. \ref{zeropvar}, the $LDoS$ of zero modes is shown for a varying distance from the wormhole bridge in the units of the radius $a$ of the wormhole center. It was also acquired in the numerical way. For the unperturbed case (0 defects), the resulting plot resembles a line. In \cite{nanommta}, the exponential solution is found for this case but with a very slow increase, so, this could be that case. It is also seen from the plot that for the increasing number of the defects the solution is approaching expressions (\ref{zero1}), (\ref{zero2}) for the zero modes of the unperturbed wormhole.

Of course, the massive fermions could also appear in the case of the perturbed wormhole. We will not perform a detailed derivation of the electronic structure for the case of this eventuality and we will only note that the corrections to the $LDoS$ would be an analogy of the corrections shown in Fig. \ref{fgMass}.\\

\subsubsection*{Graphene blackhole}

In \cite{atanasov}, the effects accompanying the deformation of the graphene and the consequent change of the distance of the carbon atoms in the layer are draught. It causes the rotation of the $p_z$ orbitals and rehybridization of the $\pi$ and $\sigma$ orbitals. This procedure leads to the creation of the $p-n$ junctions similarly to the case of a transistor. This effect changes the Fermi level which is rising in the far areas from the wormhole center. As a result, the electron flux is directed from these areas to the middle where the electric charge is accumulated. In the case of the deformed wormhole we speak about the so-called graphene blackhole. The form of the nanotube in the middle plays a big role for this purpose. It cannot be unperturbed because in such a case the effect of the blackhole would be disrupted. It can be ensured only in the case when the nanotubular neck is tapering in the direction to its center, because this ensures the decrease of the Fermi level \cite{pds, pds2, pds3, pdsf}. The related effects appearing on the nanostructures are also described in \cite{beltrami}. Here, the special relativistic-like properties of the Beltrami pseudosphere naturally point to quantum field theory in curved space. It predicts the finite temperature local density of states that is a realization of the Hawking-Unruh effect.

The effect of the graphene blackhole could eventually disappear in the presence of the external magnetic (electric) field which would cause the transfer of the charge from one of the wormhole sheets to another through the center. This serves as an important model for further investigations of the electron flux in the presence of the defects. In \cite{charge}, some investigations were carried out for the above mentioned wormhole with $12$ heptagonal defects. Possible investigations in the case of the next deformations could contribute to the applications in the cosmological models.\\

\subsection{The smallest wormhole}\

\begin{figure}[htbp]
\includegraphics[width=160mm]{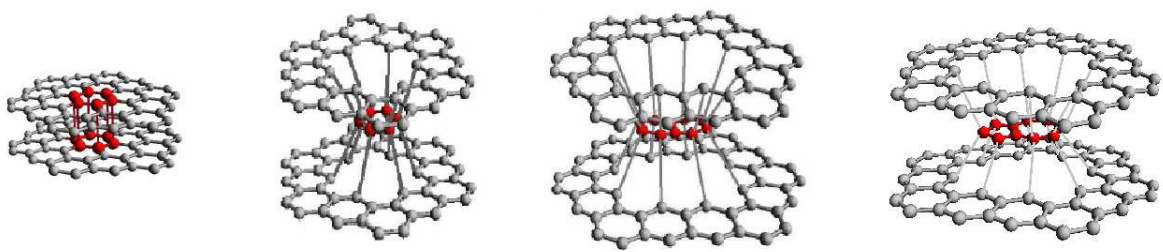}
\caption{{\fontfamily{phv}\selectfont {\fontsize{11}{0}\selectfont \textbf{Drafts of different kinds of the smallest wormhole.}}}}\label{MinW}
\end{figure}

For the purpose of the synthesis of the simplest forms of the graphene (perturbed or unperturbed) wormhole, we establish the model of the smallest wormhole where DFT and Ab-initio
computations could be developed. Possible examples of these simplest forms are shown in Fig. \ref{MinW}. Let us notice that in the connection points of the monolayers, the $sp^2$ hybridization is broken and it is replaced by the $sp^3$ hybridization but we still could detect the zero modes as the results of the $sp^2$ hybridization on the wormhole bridge. We can also imagine the presented structures in the way that the middle part is contained in the upper or the lower monolayer and both monolayers are connected through this part.\\

For some of the DFT and Ab-initio computations on graphene see, e.g., \cite{abinitio1, abinitio2}. The geometry in the case of the smallest wormhole may be equivalent to the geometry of the catenoid \cite{saxena1, katenoid}: it is
\begin{equation}{\rm d}s^2=\frac{((\eta^{\pm}\pm 1)^2+1)^2}{4(\eta^{\pm}\pm 1)^4}({\rm d}\eta^{\pm})^2+\frac{1}{4}(\frac{(\eta^{\pm}\pm 1)^2+1}{\eta^{\pm}\pm 1})^2({\rm d}\phi)^2,\end{equation}
where
\begin{equation}\eta^+=e^{\zeta_+}-1,\,\zeta_+>0,\hspace{7mm}\eta^-=-(e^{-\zeta_-}-1),\,\zeta_-<0.\end{equation}
Then, using the coordinate $\zeta$, the metric tensor could be written as
\begin{equation}\label{metric_sw}g_{\mu\nu}=\cosh^2\zeta_{\pm}\left(\begin{array}{cc}1 & 0\\0 & 1\end{array}\right),
\end{equation}
and the radius vector has the form
\begin{equation}\vec{R}(\zeta_{\pm}, \varphi)=(\cosh\zeta_{\pm}\cos\varphi,\cosh\zeta_{\pm}\sin\varphi,\zeta_{\pm}).\end{equation}
The coordinates $\eta^+, \eta^-$ play a similar role as the polar coordinates $r_+, r_-$ in the case of the unperturbed wormhole. With the help of these coordinates, one covers the entire manifold with two coordinate patches. One patch covers the region $\zeta >0$ and the other one $\zeta <0$, the upper and lower graphene layer, respectively.  The coordinates should in particular result in a metric which is reminiscent of polar coordinates at infinity and gives rise to the Bessel equation in the asymptotic region on the catenoid which is exactly the solution of the zero modes for the non-perturbed wormhole bridge. With such new coordinates we can find zero mode solutions separately on two bridges of the graphene bilayer wormhole.\\

\begin{figure}[htbp]
\includegraphics[width=85mm]{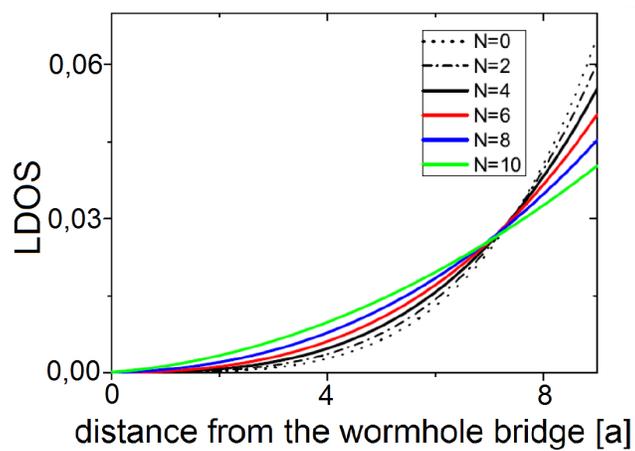}
\caption{{\fontfamily{phv}\selectfont {\fontsize{11}{0}\selectfont \textbf{Zero modes for the smallest wormhole.}}}}\label{LDsw}
\end{figure}

The zero modes are the solution of the system of equations which we get using (\ref{system}):
\begin{equation}\frac{\partial_{\zeta}u_j}{\sqrt{g_{\zeta\zeta}}}-\frac{\widetilde{j}}
{\sqrt{g_{\varphi\varphi}}}u_j
=0,\hspace{1cm}-\frac{\partial_{\zeta}v_j}{\sqrt{g_{\zeta\zeta}}}-\frac{\widetilde{j}}
{\sqrt{g_{\varphi\varphi}}}v_j
=0.\end{equation}
Since the metric coefficients in (\ref{metric_sw}) have the same form for both possibilities $\zeta_+, \zeta_-$, the variable $\zeta$ in the last system includes also both possibilities. Using the substitution $\cosh\zeta=r$, the system is transformed into
\begin{equation}\partial_{r}u_j-\frac{\tilde{j}}{\sqrt{r^2-1}}u_j
=0,\hspace{1cm}-\partial_{r}v_j-\frac{\tilde{j}}{\sqrt{r^2-1}}v_j
=0\end{equation}
and  we get the solution
\begin{equation}u_j=(r+\sqrt{r^2-1})^{\tilde{j}},\hspace{0.5cm}
v_j=(r-\sqrt{r^2-1})^{\tilde{j}}\end{equation}
and the $LDoS$ has the form
\begin{equation}LDoS(r)=C\cdot\frac{1}{r}\left[(r+\sqrt{r^2-1})^{\tilde{j}}+(r-\sqrt{r^2-1})^{\tilde{j}}\right],\end{equation}
where $C$ is the normalization constant. For the different numbers of the defects, the dependence of zero modes on the distance from the smallest wormhole center see Fig. \ref{LDsw}. Because of the symmetry of the catenoid, we get an analogous solution we get for $\zeta<0$.\\

If we compare the calculated zero modes for the smallest wormhole with the zero modes for the perturbed wormhole in Fig. \ref{zeropvar}, we see a significant difference, but one of the features is the same: the more defects we have, the more significant the corresponding zero mode close to the wormhole bridge is. We can suppose that the more the shape of the smallest wormhole approaches the shape of the perturbed wormhole, the better agreement of both Figures we get.\\

\section{Synthesis of the draught materials}\

\begin{figure}[htbp]
\includegraphics[width=80mm]{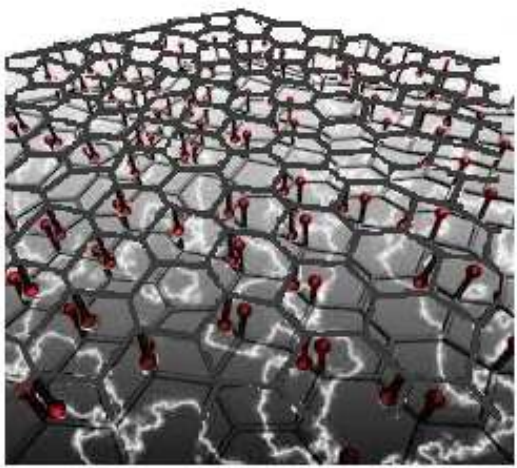}
\caption{{\fontfamily{phv}\selectfont {\fontsize{11}{0}\selectfont \textbf{Doping the graphene bilayer with the reactive molecules.}}}}\label{fg_dope}
\end{figure}

The ways how to synthesize the described material are still at the theoretical stage.
We predict that it could be manufactured from the graphene bilayers whose properties are described, for example, in \cite{mucha} or \cite{jernigan}. We consider that the graphene monolayers could be mechanically pressed against each other so that their distance would be reduced below the value of the length of the atomic bonds in the graphene. Under these conditions, the interaction between the valence electrons of the carbon atoms from the opposite layers could achieve significant values, because it would exceed the interaction between the neighbors in the hexagonal carbon structure. Furthermore, as we considered above, the radius of the wormhole must be much smaller than the graphene layers length, so the minimal distance between the monolayers is very important. The structure of the wormhole could then arise spontaneously. This should be a concrete task for future experimental research.
Next of the possibilities is doping the graphene sheets with the reactive molecules (see Fig. \ref{fg_dope}) and then pressing the sheets against each other. The connections between the sheets, which would be formed by the wormholes, would be created by the mentioned molecules. The places where the wormholes would be present could be determined by the detection of the zero modes.\\

\section{Conclusions}\

The wormhole allotropes could be prepared by the procedure which would be composed of two steps: first, doping the graphene sheets by the reactive molecules and second, pressing these sheets against each other by achieving their connection through the mentioned molecules. The graphitic wormhole could also be created spontaneously in the bilayer graphene. The created wormholes in the graphene structure could then be indicated by the presence of the corresponding zero modes. We use the continuum gauge field-theory for their calculation. It was shown in the calculations of the $LDoS$ for the perturbed wormhole that the closer the number of the defects to the border values 0 a 12 is, the more the $LDoS$ coincides with the calculations carried out for these 2 values, as can be verified by the comparison with \cite{herrero1} and \cite{nanommta}.

Next possibilities as to identify the wormhole structure in the graphene bilayer is via spin-orbit interaction or coupling (SOC) in connecting nanotubes \cite{soctube}. The SOC in graphene could be induced as well by the nonzero curvature \cite{soc,soc1,soc2,soc3}. In the case of the perturbed wormhole with negative curvature the chiral fermions penetrating through the connecting nanotube in the wormhole structure could be created. The higher this effect is, the lower the radius of the connecting wormhole is. Moreover, the chiral fermions should prefer permanently only one direction of the massive or massless fermionic current \cite{fercur}, e.g., from the upper graphene sheet to the lower in the graphene bilayer through the connecting nanotube depending on the wormhole curvature. This permanently oriented flow could be detected by the experimental observations.

There is also a real possibility that the wormhole bridge could
serve as some trap for quantum dots \cite{dots} or some hydrogen
impurities in graphene \cite{hydrog}. The wormhole bridge for their
special massive chiral fermions could also absorb very effectively all
incident light of a specific wavelength coming from any direction,
similarly described in\cite{dots}.\\

\begin{figure}[htbp]
\includegraphics[width=70mm]{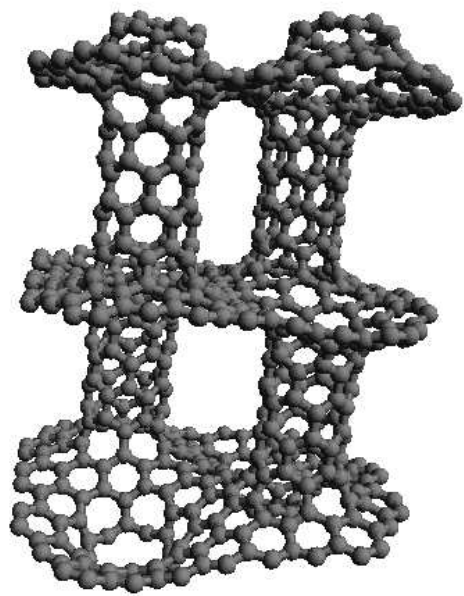}
\caption{{\fontfamily{phv}\selectfont {\fontsize{11}{0}\selectfont \textbf{Fragment of pillared graphene.}}}}\label{fgPill}
\end{figure}

More complicated wormhole structures can be established. One of them is the pillared graphene (Fig. \ref{fgPill}). This kind of nanostructures will have a wide application in the branch of storage of the hydrogen fuel \cite{dmitr1}. Initially, the carbon nanotubes were considered as a suitable candidate, but for different reasons, e.g. the insufficiently large molecular surface area, it was shown that these structures are not suitable and the pillared graphene could be much better for this purpose. One of the possible methods of the synthesis of this kind of nanostructure was described in \cite{pillful}. Here the procedure is suggested in which the layer of the fullerene molecules is inserted between the graphene sheets and the reactive molecules are put to the attachment points. Then, the required chemical structure is achieved via the thermal and radiative processes.\\

\begin{figure}[htbp]
\includegraphics[width=150mm]{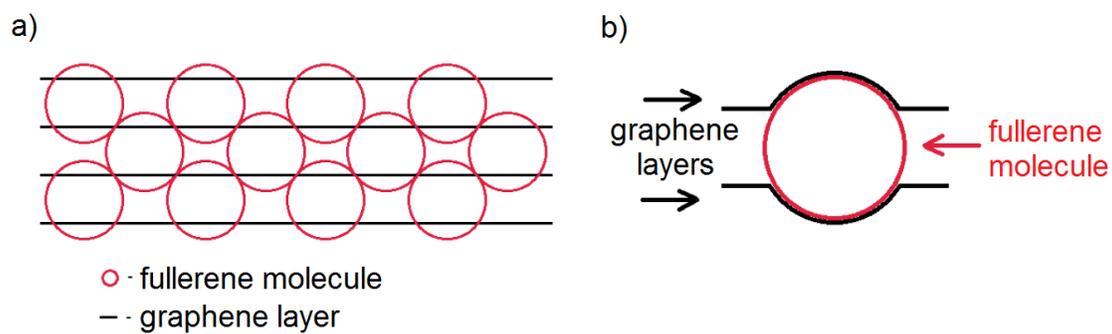}
\caption{{\fontfamily{phv}\selectfont {\fontsize{11}{0}\selectfont \textbf{The configuration of the fullerene molecules between the graphene layers in the process of the synthesis of the pillared graphene.}}}}\label{ful_lay}
\end{figure}

\begin{figure}[htbp]
\includegraphics[width=150mm]{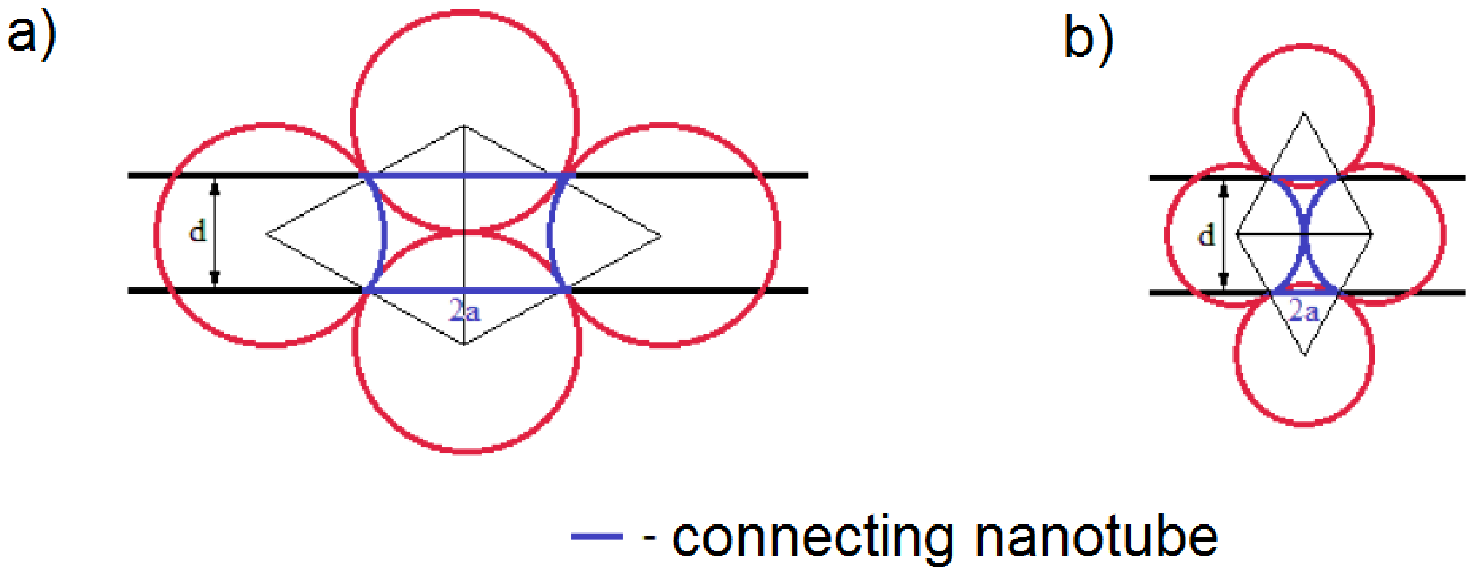}
\caption{{\fontfamily{phv}\selectfont {\fontsize{11}{0}\selectfont \textbf{The extremal sizes of the ratio $d/2a$ in the connecting nanotube: a) minimal, b) the maximal.}}}}\label{ekstrem}
\end{figure}

The density of the fullerene molecules in the corresponding layer could influence the size of the connecting nanotubes in the pillared graphene. This can be seen in Fig. \ref{ful_lay}a. There, the particular fullerene layers, which deform the graphene layers, fit to each other tightly. This is only the schematic sketch, the deformation of the graphene layers by the fullerene molecules is shown in Fig. \ref{ful_lay}b. Each of the connecting nanotubes arises in the place of the contact points. The ratio of the hight and the width of the connecting nanotube has the value between 2 extremal values whose origin is illustrated in Fig. \ref{ekstrem}. It is possible to easily derive that in the case a), this ratio has the value $1/\sqrt{3}\doteq 0.577$ and in the case b) the ratio is $\sqrt{3}/2\doteq 0.866$. So, in the general case, we have for the size of the connecting nanotube,
\begin{equation}0.577\,\leq\,d/2a\,\leq\,0.866.\end{equation}

Finally, the question arises if we could design condensed-matter systems to test the supposedly non-testable predictions of string theory too? Another question to ask, though, is whether what we think of as the fundamental laws of physics, such as quantum field theory, themselves emerge from some complex inner structure that remains inaccessible to us.

Recently, it was found in work \cite{letter} that the electron induced rippling in the graphene must be unstable towards a buckling transition that is the analogue of Higgs condensation, showing another way to employ graphene as a test ground of fundamental concepts in theoretical physics. The graphitic wormhole structure for their extreme curvature and a very thin connecting nanotube is the best candidate for such experiments to test also string theory as high energy physics peculiarities.

Summarize, the presented paper is a real connection between nanoscale physics, condensed matter physics, chemical physics, computational physics, relativistic physics and experimental physics. We hope that it reflects a lot of main ideas of the $21^{st}$ century of scientific research. Similar topics are described for example in the Refs. \cite{kren, zimb, RooKhad, rodrig, paoloDS, YNL, KKMR, RWang}.\\

ACKNOWLEDGEMENTS --- The authors thank Dr. M. Pudlak for helpful discussions and advice especially in the spin-orbit coupling part of work.

The work was supported
by the Slovak Academy of Sciences in the framework of CEX NANOFLUID,
and by the Science and Technology Assistance Agency under Contract
No. APVV-0509-07 and APVV-0171-10, VEGA Grant No. 2/0037/13 and Ministry
of Education Agency for Structural Funds of EU in frame of project
26220120021, 26220120033 and 26110230061. R. Pincak would like to thank the
TH division in CERN for hospitality.

\newpage

\newpage

\newpage

\newpage

\newpage

\newpage

\newpage

\newpage

\newpage

\newpage

\newpage

\newpage

\newpage

\newpage

\newpage

\newpage

\newpage

\begin{thebibliography}{99}

\bibitem{herrero1}
J. Gonzalez, F. Guinea and J. Herrero, \textit{Phys. Rev. B 79}, 165434 \textbf{(2009)}.

\bibitem{herrero2}
J. Gonzalez and J. Herrero, \textit{Nucl. Phys. B 825}, 426443 \textbf{(2010)}.

\bibitem{tuzun}
B. T\"{u}z\"{u}n and C. Erko\c{c}, \textit{Quantum Matter 1}, 136 \textbf{(2012)}.

\bibitem{fiscal}
D. Fiscaletti, \textit{Quantum Matter 2}, 45-53 \textbf{(2013)}.

\bibitem{cone}
M. Sanderson, Y. S. Ang and C. Zhang, \textit{Phys. Rev. B Condens. Matter 88 (24)}, 245404-1-245404-9 \textbf{(2013)}.

\bibitem{pinw}
R. Pincak and J. Smotlacha, \textit{Eur. Phys. J. B 86}, 480 \textbf{(2013)}.


\bibitem{mele}
D. P. DiVincenzo and E. J. Mele, \textit{Phys. Rev. B 29}, 1685 \textbf{(1984)}.


\bibitem{pinbil}
M. Pudlak and R. Pincak, \textit{Eur. Phys. J. B 86}, 107 \textbf{(2013)}.

\bibitem{bilayer}
E. V. Castro, M. P. Lopez-Sancho and M. A. H. Vozmediano, \textit{Phys. Rev. Lett. 104}, 036802 \textbf{(2010)}.

\bibitem{trilayer}
E. V. Castro, M. P. Lopez-Sancho and M. A. H. Vozmediano, \textit{Solid State Commun. 152}, 1483 \textbf{(2012)}.

\bibitem{pin}
E. A. Kochetov, V. A. Osipov and R. Pincak, \textit{J. Phys.: Condens. Matter 22}, 395502 \textbf{(2010)}.

\bibitem{vf}
H. Rostami and R. Asgari, \textit{Phys. Rev. B 86}, 155435 \textbf{(2012)}.

\bibitem{chenwei}
Chen-Wei Jiang, Xiang Zhou, Rui-Hua Xie and Fu-Li Li, \textit{Quantum Matter 2}, 353-363 \textbf{(2013)}.

\bibitem{blvf0}
G. Borghi, M. Polini, R. Asgari and A. H. MacDonald, \textit{Solid State Commun. 149}, 1117 \textbf{(2009)}.

\bibitem{blvf}
M. I. Katsnelson, K. S. Novoselov and A. K. Geim, \textit{Nature Phys. 2}, 620 \textbf{(2006)}.

\bibitem{tubmass}
A. D. Alhaidari, A. Jellal, E. B. Choubabi and H. Bahlouli, \textit{Quantum Matter 2}, 140 \textbf{(2013)}.

\bibitem{ten}
W. Greiner, \textit{Relativistic Quantum Mechanics: Wave Equations}, Springer, Berlin
\textbf{(1994)}; B. Thaller, \textit{The Dirac Equation}, Springer, Berlin \textbf{(1992)}.

\bibitem{enerbulk}
P. H. Tan, \textit{Nature Mater. 11}, 294 \textbf{(2012)}.

\bibitem{soliton}
J. S. Alden, A. W. Tsen, P. Y. Huang, R. Hovden, L. Brown, J. Park, D. A. Muller and P. L. McEuen, \textit{http://arxiv.org/abs/1304.7549}, \textbf{(2013)}.

\bibitem{spoc1}
C. L. Kane and E. J. Mele, \textit{Phys. Rev. Lett. 95}, 226801 \textbf{(2005)}.

\bibitem{spoc2}
T. Ando, \textit{J. Phys. Soc. Jpn. 69}, 1757 \textbf{(2000)}.

\bibitem{nanommta}
R. Pincak, J. Smotlacha and M. Pudlak, \textit{NanoMMTA 2}, 81 \textbf{(2013)}.

\bibitem{atanasov}
V. Atanasov and A. Saxena, \textit{Phys. Rev. B 81}, 205409 \textbf{(2010)}.

\bibitem{pds}
M. Pudlak and R. Pincak, \textit{Eur. Phys. J. B 67}, 565 \textbf{(2009)}.

\bibitem{pds2}
M. Pudlak and R. Pincak, \textit{Phys. Rev. A 79}, 033202 \textbf{(2009)}.

\bibitem{pds3}
R. Pincak, M. Pudlak and J. Smotlacha, in \textit{Carbon Nanotubes:Synthesis, Properties and Applications}, NOVA Science Publisher, New York \textbf{(2012)}.

\bibitem{pdsf}
R. Pincak and M. Pudlak, in \textit{Progress in Fullerene Research}, Edited F. Columbus, NOVA Science Publisher, New York \textbf{(2007)}.

\bibitem{beltrami}
A. Iorio and G. Lambiase, \textit{Phys. Lett. B 716}, 334 \textbf{(2012)}.

\bibitem{charge}
E. Guendelman, A. Kaganovich, E. Nissimov and S. Pacheva, \textit{Open Nucl. Part. Phys. J.  4}, 27 \textbf{(2011)}.

\bibitem{abinitio1}
H. Min, B. Savu, S. K. Banerjee and A. H. MacDonald, \textit{Phys. Rev. B 75}, 155115 \textbf{(2007)}.

\bibitem{abinitio2}
D. Solenov, C. Junkermeier, T. L. Reinecke and K. A. Velizhanin, \textit{Phys. Rev. Lett. 111}, 115502 \textbf{(2013)}.

\bibitem{saxena1}
R. Dandoloff, A. Saxena and B. Jensen, \textit{Phys. Lett. A 373}, 2667 \textbf{(2009)}.

\bibitem{katenoid}
V. Atanasov and A. Saxena, \textit{J. Phys.: Condens. Matter 23}, 175301 \textbf{(2011)}.

\bibitem{mucha}
M.Mucha-Kruczy�ski, I. L. Aleiner and V. I. Fal�ko, \textit{Phys. Rev. B 84}, 041404(R) \textbf{(2011)}.

\bibitem{jernigan}
G. G. Jernigan, \textit{J. Vac. Sci. Technol. B 30}, 03D110 \textbf{(2012)}.

\bibitem{soctube}
F. Kuemmeth, S. Ilani, D. C. Ralph and P. L. McEuen, \textit{Nature 452}, 448 \textbf{(2008)}.

\bibitem{soc}
M. V. Entin and L. I. Magaril, \textit{Phys. Rev. B 64}, 085330 \textbf{(2001)}.

\bibitem{soc1}
A. De Martino, R. Egger, K. Hallberg and C. A.  Balseiro, \textit{Phys. Rev. Lett. 88}, 206402 \textbf{(2002)}.

\bibitem{soc2}
J. S. Jeong and H. W. Lee, \textit{Phys. Rev. B 80}, 075409 \textbf{(2009)}.

\bibitem{soc3}
M. P. Lopez-Sancho and M. C. Munoz, \textit{Phys. Rev. B 83}, 075406 \textbf{(2011)}.

\bibitem{fercur}
S. Bellucci, A. A. Saharian and V. M. Bardeghyan, \textit{Phys. Rev. D 82}, 065011 \textbf{(2010)}.


\bibitem{dots}
S. Thongrattanasiri, F. H. L. Koppens and F. J. G. de Abajo, \textit{Phys. Rev. Lett. 108}, 047401 \textbf{(2012)}.

\bibitem{hydrog}
G. Forte, \textit{Phys. Lett. A  372}, 6168 \textbf{(2008)}.

\bibitem{dmitr1}
G. Dmitrakakis, \textit{SPIE newsroom}, \textbf{(2009)}.



\bibitem{pillful}
D. L. Barker, W. R. Owens and J. W. Beck, Fabrication of pillared graphene. U.S. Patent 20120152725, Jun 21 \textbf{(2012)}.

\bibitem{letter}
P. San-Jose, J. Gonzalez and F. Guinea, \textit{Phys. Rev. Lett. 106}, 045502 \textbf{(2011)}.



\bibitem{kren}
A. Khrennikov, \textit{Rev. Theor. Sci. 1}, 34-57 \textbf{(2013)}.

\bibitem{zimb}
N. A. Zimbovskaya, \textit{Rev. Theor. Sci. 2}, 1-45 \textbf{(2014)}.

\bibitem{RooKhad}
A. Roohi and H. Khademolhosseini, \textit{Rev. Theor. Sci. 2}, 46-76 \textbf{(2014)}.

\bibitem{rodrig}
C. B. Rodrigues dos Santos, C. C. Lobato, M. A. Costa de Sousa, W. J. da Cruz Macedo, and J. C. T. Carvalho,
\textit{Rev. Theor. Sci. 2}, 91-115 \textbf{(2014)}.

\bibitem{paoloDS}
P. Di Sia, \textit{Rev. Theor. Sci. 2}, 146-180 \textbf{(2014)}.

\bibitem{YNL}
J. J. Yeo, T. Y. Ng, and Z. S. Liu, \textit{J. Comput. Theor. Nanosci. 11}, 1790-1796 \textbf{(2014)}.

\bibitem{KKMR}
F. Kiani, T. Khosravi, F. Moradi, P. Rahbari, M. J. Aghaei, M. Arabi, H. Tajik, and R. Kalantarinejad,
\textit{J. Comput. Theor. Nanosci. 11}, 1237-1243 \textbf{(2014)}.

\bibitem{RWang}
B. M. Raffah and J. Wang, \textit{J. Comput. Theor. Nanosci. 11}, 1049-1054 \textbf{(2014)}.

\end{thebibliography}
\end{document}